\documentclass[aps,pra,superscriptaddress]{revtex4-2}

\usepackage{graphicx}
\usepackage{dcolumn}
\usepackage{bm}
\usepackage{braket}
\usepackage[utf8]{inputenc}
\usepackage[T1]{fontenc}
\usepackage{mathptmx}
\usepackage{siunitx}
\usepackage{url}
\usepackage{subcaption}
\usepackage{hyperref}
\usepackage[dvipsnames]{xcolor}
\usepackage{ulem}
\usepackage{siunitx}
\usepackage[export]{adjustbox}
\usepackage{svg}
\usepackage{booktabs}
\usepackage{tabularx}
\usepackage{anyfontsize}

\newcolumntype{Y}{>{\centering\arraybackslash}X}
\newcolumntype{L}[1]{>{\hsize=#1\hsize\raggedright\arraybackslash}X}%
\newcolumntype{R}[1]{>{\hsize=#1\hsize\raggedleft\arraybackslash}X}%
\newcolumntype{C}[1]{>{\hsize=#1\hsize\centering\arraybackslash}X}%

\begin{document}


\title{3D Printed Monolithic Optomechanical Chassis for an Unbalanced Field-Widened Interferometer}



\author{Ramy Tannous}
\email{ramy.tannous@uwaterloo.ca}
\altaffiliation[Current address:]{ National Research Council of Canada, 100 Sussex Drive, Ottawa ON, Canada}
\affiliation{Institute for Quantum Computing, University of Waterloo, 200 University Avenue W, Waterloo, ON N2L3G1, Canada}
\affiliation{Department of Physics and Astronomy, University of Waterloo, 200 University Avenue W, Waterloo, ON N2L3G1, Canada}

\author{Dogan Sinar}
\email{dsinar@uwaterloo.ca}
\affiliation{Institute for Quantum Computing, University of Waterloo, 200 University Avenue W, Waterloo, ON N2L3G1, Canada}
\affiliation{Department of Physics and Astronomy, University of Waterloo, 200 University Avenue W, Waterloo, ON N2L3G1, Canada}

\author{Sagar Patel}
\affiliation{Department of Mechanical and Mechatronics Engineering, University of Waterloo, Waterloo, ON, N2L 3G1 Canada}

\author{Tabitha D. Arulpragasam}
\affiliation{Institute for Quantum Computing, University of Waterloo, 200 University Avenue W, Waterloo, ON N2L3G1, Canada}
\affiliation{Department of Chemical Engineering, University of Waterloo, 200 University Avenue W, Waterloo, ON N2L3G1, Canada}

\author{Issa Rishmawi}
\affiliation{Department of Mechanical and Mechatronics Engineering, University of Waterloo, Waterloo, ON, N2L 3G1 Canada}

\author{Mark Kirby}
\affiliation{Department of Mechanical and Mechatronics Engineering, University of Waterloo, Waterloo, ON, N2L 3G1 Canada}

\author{Vlad Paserin}
\affiliation{Department of Mechanical and Mechatronics Engineering, University of Waterloo, Waterloo, ON, N2L 3G1 Canada}

\author{Mihaela Vlasea}
\affiliation{Department of Mechanical and Mechatronics Engineering, University of Waterloo, Waterloo, ON, N2L 3G1 Canada}

\author{Thomas Jennewein}
\email{thomas.jennewein@uwaterloo.ca}
\affiliation{Institute for Quantum Computing, University of Waterloo, 200 University Avenue W, Waterloo, ON N2L3G1, Canada}
\affiliation{Department of Physics and Astronomy, University of Waterloo, 200 University Avenue W, Waterloo, ON N2L3G1, Canada}
\affiliation{Department of Physics, Simon Fraser University, 8888 University Dr W, Burnaby, BC V5A 1S6, Canada}


\date{\today}

\begin{abstract}
Additive manufacturing is increasingly being explored as an approach for producing complex structures for optical hardware. In this work, we present a fully monolithic optomechanical chassis fabricated using laser powder bed fusion additive manufacturing. The structure was fabricated from thermally stable Invar36, a low–thermal‑expansion alloy that enables high dimensional stability under varying environmental conditions. Our monolithic design incorporates embedded optical mounts, flexures, and alignment features directly into the chassis geometry that exceed the practical limits of conventional subtractive manufacturing. By integrating the optomechanical support structure into a single compact unit, the system achieves a reduced footprint compared to traditional optomechanical assemblies. The results demonstrate the potential of additive manufacturing to produce lightweight, and highly customized optical systems, opening the possibility of new pathways for next‑generation quantum sensors and optical instrumentation.
\end{abstract}


\maketitle
\section{Introduction}
\label{sec:intro}

Demand for complex yet robust and practical optical systems has been increasing, particularly with the emergence of quantum computing and sensing technologies~\cite{cohen2019thresholded, haase202132x24, liu2020inspiring}. Many optical applications demand optomechanical enclosures that integrate precision alignment and readjustment mechanisms, yet still preserve a modular and reconfigurable system layout. Commercial optomechanical components are typically designed for general‑purpose use and often fail to meet the constraints of compact or highly integrated optical systems. Additive manufacturing (AM)~\cite{beaman2020additive} enables rapid fabrication of application‑specific geometries, allowing designers to tailor mechanical and optical features directly to system requirements. Moreover, AM supports the creation of mounts, flexures~\cite{chun2020review_flexure_withAdditive}, and lightweight structural elements~\cite{DUPLESSIS2022100918_metal_AM_devices} that cannot be produced through conventional machining, facilitating both mass‑reduction strategies and the incorporation of custom features optimized for the optical components in use.

Published literature on compact and sophisticated optomechanical designs using AM is scarce. Several studies have been conducted over the last decade, investigating the potential use of AM for optomechanical components, particularly discreet optical mounts and flexures manufactured using polymer filament 3D printers~\cite{salazar20173dtoolbox,winters20183dCageSystem,zhang2013openSource}. A common theme of these studies is the use of traditional optomechanical designs for production through AM. Consequently, several components had to be manufactured in pieces and later assembled ~\cite{hussain2018designPhone,kranert20203dLowCost,mammini20153dOpticalMount}.
A quasi-monolithic system was demonstrated where a low power laser system was 3D printed using polymers as a single piece except for the adjustable pump mirror, which was assembled post-printing~\cite{kranert2021Quasi}. Additive manufacturing has been used for quantum technologies, particularly a ultra-high vacuum chamber~\cite{COOPER2021101898_vaccuum} and a fully monolithic interferometer that was used for a trapped atom system was demonstrated~\cite{madkhaly2021performance}. However, this monolithic system did not have any optomechanical alignment components in the monolithic portion of the interferometer. 

To the knowledge of the authors, this work represents the first demonstration of a fully 3D printed monolithic optomechanical enclosure for an interferometer with built-in adjustable flexure mounts. The enclosure houses a field‑widened interferometer capable of interfering multimode optical signals, a configuration particularly relevant for quantum information applications~\cite{jin2018demonstration,tannous2025towards,vinet2026time}. We use laser bed powder fusion (LBF/PBF-LB) as the additive manufacturing technique to print our prototype from Invar36, a low thermal expansion metal. The use of Invar36 reduces the thermal sensitivity of the system, making it less prone to phase fluctuations. Although additive manufacturing has been demonstrated to produce tools and equipment comparable to commercial alternatives, conventional manufacturing is capable at producing highly reliable tools and equipment in near perfect form. Hence, in the scope of this study, any optomechanical design should take advantage of the specific benefits offered by additive manufacturing. With these considerations in mind, we introduce a monolithic optical enclosure that integrates all alignment and mounting features directly into the printed structure. Commercial optomechanical components are typically designed for general‑purpose use and are often difficult to incorporate into compact interferometer layouts due to their bulky form factors. By contrast, our monolithic chassis enables a custom optomechanical architecture with a reduced physical footprint, embedded alignment mechanisms, and the potential for improved mechanical and thermal stability. 



\section{Additive Manufacturing}
\label{sec:additive}
 Metal AM technologies such as laser powder bed fusion (LPBF) and electron beam powder bed fusion (EB-PBF), also known as electron beam melting, are the leading metal AM technologies used for industrial part production for aerospace~\cite{DebRoy2019Scientific, 3DPrinti82:online}, biomedical~\cite{mcgregor2021architectural, Worldfir36:online}, and automotive~\cite{Fraunhof28:online, brock2021relative} applications. Amongst all available metal AM technologies, LPBF has the highest industrial uptake at the moment~\cite{khorasani2020review}. The main reasons for the advancement of LPBF involve: a) the finest feature resolution possible due to the beam spot sizes generally being between 50-100 \textmu m~\cite{patel2020melting}, b) a wide range of materials compatible with the technology~\cite{sing2016laser}, c) the advent of design for AM which open up the potential to high-complex geometries~\cite{thompson2016design}, and d) the potential to obtain performance superior to conventional manufacturing processes~\cite{wang2018additively}. 

LPBF is the AM technology selected for this study. The process involves the spreading of thin layers of micron sized metal powder sequentially over a metallic build-plate. A high-power laser beam is then scanned over each layer at locations specified by the input computer aided design (CAD) file. The laser selectively melts and fuses the powder particles together according to the cross-sectional geometry of the part for that specific layer. Due to the concentrated energy input of the laser beam, localized thermal loads are generated, leading to numerous challenges while adopting this technology for new materials, including the material selected for this work – Invar36. At a composition of approximately 36 wt.\% Ni, the Invar36 alloy takes a characteristically low coefficient of thermal expansion (CTE) compared to most known Invar alloys~\cite{saito1978physics}, while also having reasonable elastic properties. Due to this characteristic, Invar36 alloy is widely used in many applications requiring high dimensional stability such as metrology, cryogenic liquid storage and transport, and optical mounting in the space industry~\cite{asgari2018thermal}. The low CTE of Invar36 also makes it suitable for this work, as housing interferometers with low CTE materials are advantageous to reduce thermal phase instabilities.


\section{Design}
\label{sec:design}


\subsection{Optical Design}
\label{sec:opticDes}
\begin{figure}[htbp]
    \centering
    \includegraphics[width=0.9\textwidth]{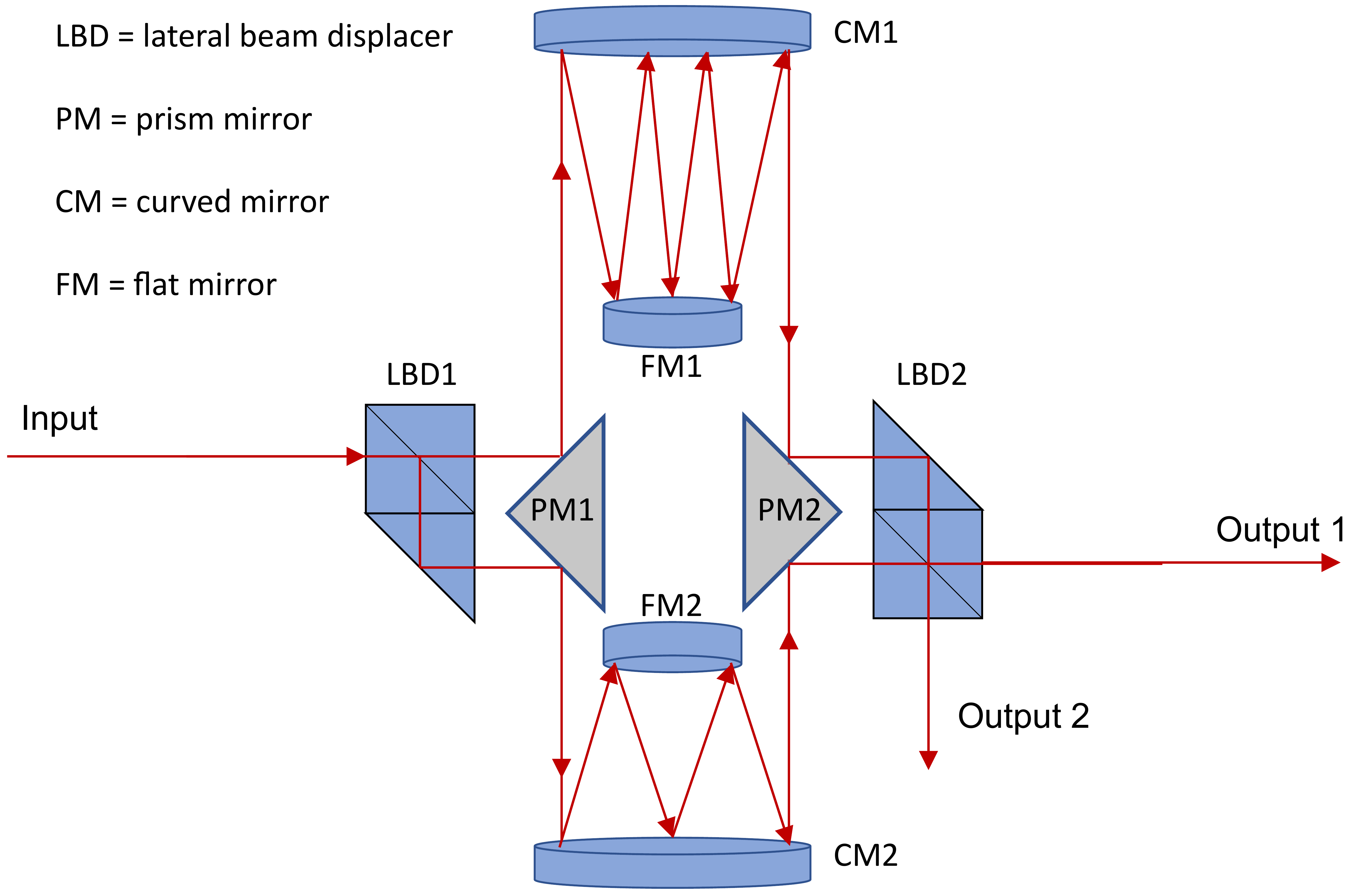}
    \caption{The conceptual optical design of a field-widened Offner relay unbalanced Mach Zehnder interferometer. }
    \label{fig:optics-setup}
\end{figure}

Before introducing the optomechanical enclosure, we first outline the optical design of the interferometer. Although optical architectures are often co‑designed with their mechanical constraints, here we treat the optical layout as the primary driver of the system requirements. The monolithic chassis developed in this work is designed to house a field‑widened, unbalanced Mach–Zehnder interferometer, whose conceptual layout is shown in Fig.~\ref{fig:optics-setup}. The field‑widening stage consists of a spherical mirror paired with a flat mirror, forming an imaging system that increases the interferometer’s field of view and enables interference of multimode input signals. By adjusting the separation between the spherical (CM) and flat mirrors (FM), the optical beam undergoes multiple reflections within the imaging system, creating a compact cavity configuration that reduces the overall physical footprint.

A detailed analysis of the field‑widened unbalanced Offner relay interferometer (ORI) and its optical performance is provided in~\cite{TannousRamy2023} and Tannous et al.~\cite{tannous2026all}. In those studies, the interferometer was implemented using conventional optomechanical components and achieved high interference visibility with multimode signals ($>98\%$). In the present work, we focus on the performance obtained when the same optical architecture is integrated into a fully monolithic, additively manufactured chassis.

\subsection{Mechanical Design}
\label{sec:optomech}
\begin{figure}[htbp]
    \centering
    \includegraphics[width=0.7\textwidth]{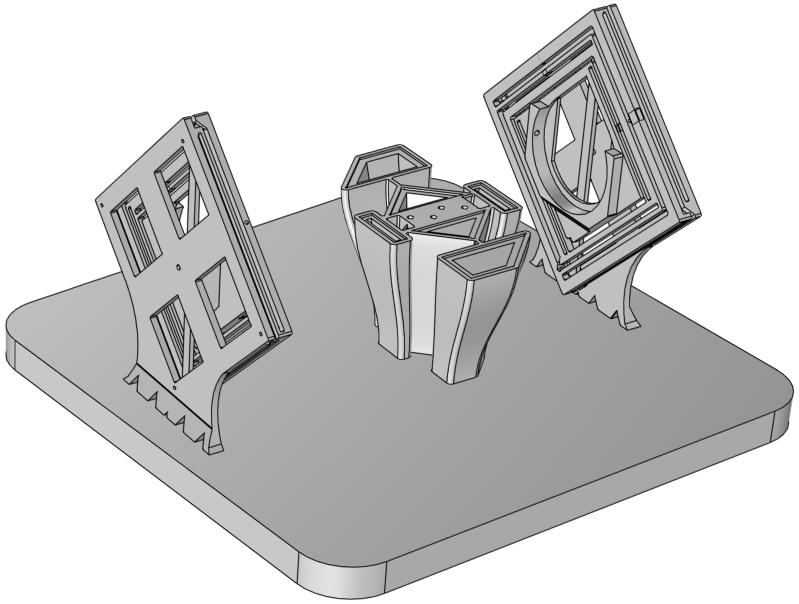}
    \caption{CAD image of the monolithic chassis. The center island and the flexures are built directly onto the rectangular LPBF build-plate.}
    \label{fig:Cadchassis}
\end{figure}

The additive manufacturing optomechanical chassis (AMOC) designed to house the ORI is shown in Fig.~\ref{fig:Cadchassis}. Throughout this manuscript, we refer to the combination of the ORI and the AMOC as the printed Offner relay interferometer (PORI). The optomechanical design of the AMOC has four goals: (1) a reduced form factor; (2) a monolithic, assembly-free structure; (3) adjustable mounts for alignment; (4) mechanical and thermal stability. (1) is achieved by the folded optical path and the monolithic design for the chassis. For this study, optimization of the placement of mounts and mount designs are not fully investigated due to use of standard off-the-self 1 and 2 inch optics. However, future work to investigate the use of smaller custom optics can help optimize the structure further reducing the form factor of the device. Smaller form factors are generally desired for resource starved platforms such as satellites~\cite{bedington2017progress} and drones~\cite{tian2024experimental}. 

\begin{figure}[htbp]
        \centering
        \begin{subfigure}[]{0.25\textwidth}
            \includegraphics[width=0.4\textwidth]{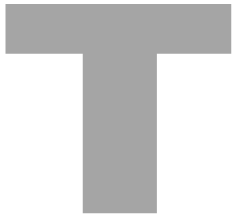}   
            \caption{}
            \label{fig:overhang_a}
        \end{subfigure}
        \begin{subfigure}[]{0.25\textwidth}
            \includegraphics[width=0.4\textwidth]{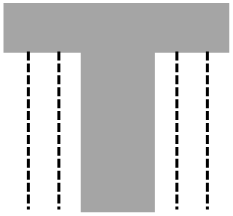}   
            \caption{}
            \label{fig:overhang_b}
        \end{subfigure}
        \begin{subfigure}[]{0.25\textwidth}
            \includegraphics[width=0.4\textwidth]{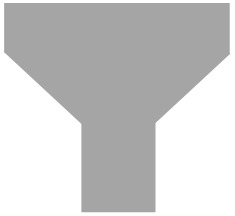}   
            \caption{}
            \label{fig:overhang_c}
        \end{subfigure}
        \caption{The overhang design in (\subref{fig:overhang_a}) can cause the built structure to collapse under its own weight. (\subref{fig:overhang_b}) Thin support structures can be printed to support the weight. (\subref{fig:overhang_c}) Support structures can be avoided all together using a \qty{45}{\degree} chamfer.}
        \label{fig:overhang}
\end{figure}

Requirement (2) can be achieved with the use of AM, provided that certain limitations are considered. As discussed in Section~\ref{sec:additive}, the LPBF AM process starts with a metal build-plate. Material is deposited onto the build-plate and is attached through a melting process. From this base, complex structures are grown vertically from the attachment point. However, an important limitation to this AM technique is "overhanging" features of a design as shown in Fig.~\ref{fig:overhang}. To avoid print failures, overhangs that cannot be supported during printing must be minimized. A common guideline is to avoid surfaces that form an angle greater than 45 degrees with the vertical, unless temporary or permanent support structures are introduced beneath them. These considerations are incorporated into the AMOC design, where only a limited number of temporary supports are required.

Proper optical simulation shows that fine positional adjustments of the spherical mirrors in each arm are sufficient to achieve the required alignment of the ORI. Because requirement (2) mandates a fully monolithic structure, conventional multi‑piece optomechanical mounts cannot be used. Instead, we employ custom monolithic flexure mounts. Unlike traditional mounts, flexures are single‑piece compliant mechanisms that provide controlled motion through elastic deformation~\cite{howell_handbook_2013}. The design of these flexures is detailed in Section~\ref{sec:flexure}, and their integration into the chassis satisfies requirement (3).

\begin{figure}[htbp]
		\centering
		\begin{subfigure}{0.65\columnwidth}
			\includegraphics[width=\columnwidth]{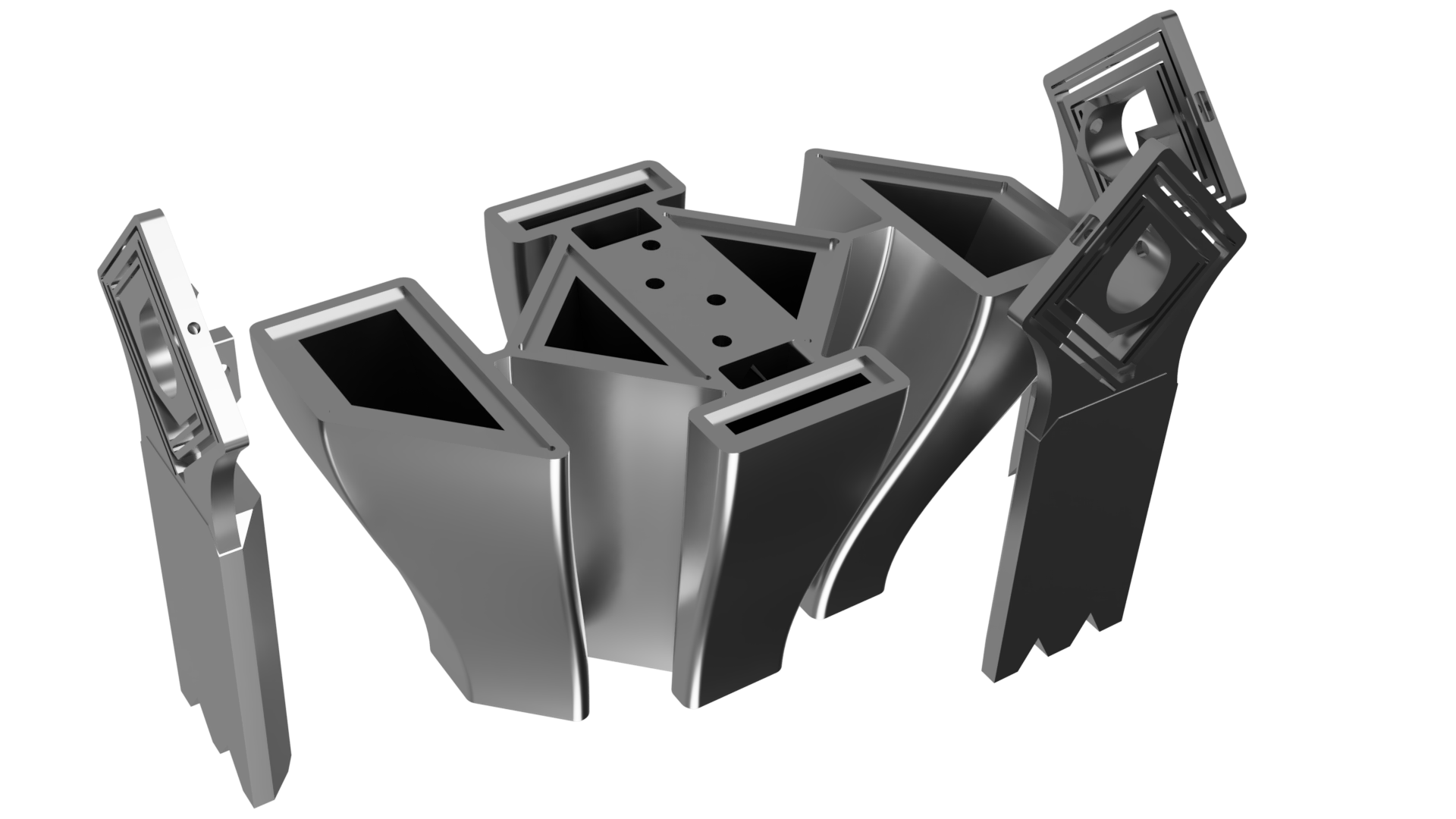}
			\caption{}
			\label{fig:loft_island_render}
		\end{subfigure}
		\\
		\begin{subfigure}{0.35\columnwidth}
			\includegraphics[width=\columnwidth]{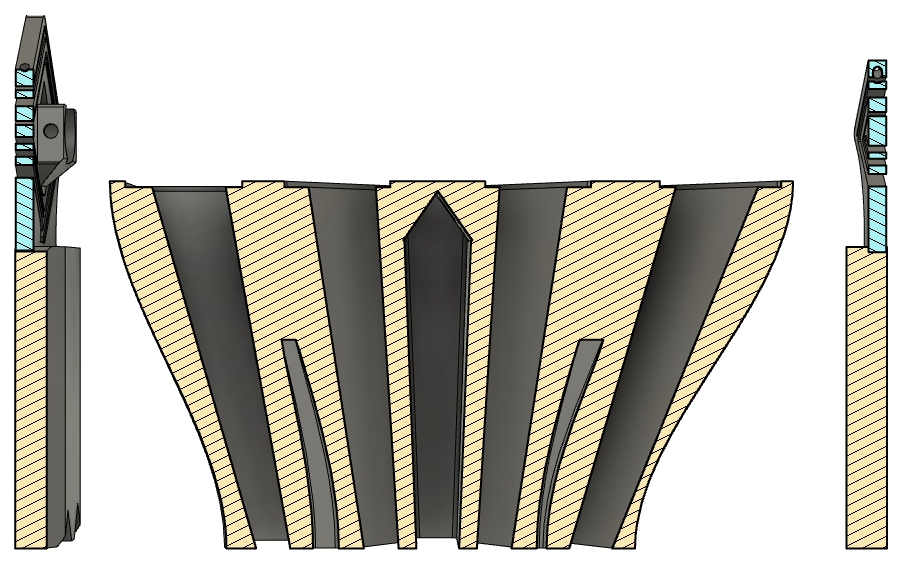}
			\caption{}
			\label{fig:loft_island_inside1}
		\end{subfigure}
		\begin{subfigure}{0.35\columnwidth}
			\centering
			\includegraphics[height=4.3cm]{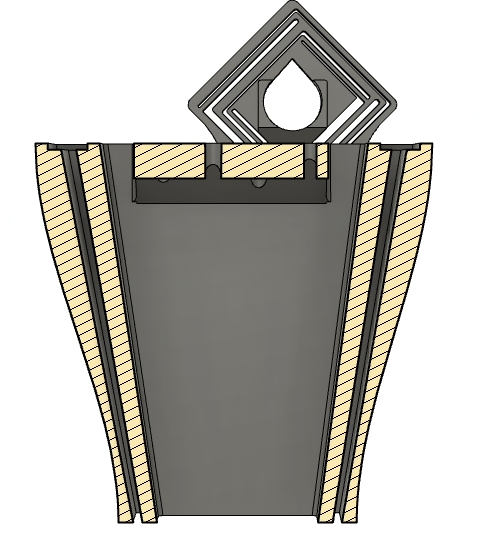}
			\caption{}
			\label{fig:loft_island_inside2}
		\end{subfigure}
		\caption[\textit{Loft Island} conceptual schematics.]{\textit{Loft Island} conceptual schematics. (\subref{fig:loft_island_render}) Render of the \textit{Loft Island} design that supports the central optics, including the input and output flexures. (\subref{fig:loft_island_inside1}) and (\subref{fig:loft_island_inside2}) Cross sectional view of the \textit{Loft Island} design showing the hollow core of each support structure.}
		\label{fig:loft_island}
	\end{figure}

In addition to the spherical mirrors, the input and output fiber‑coupling stages of the ORI are also equipped with flexures, enabling lateral adjustments in the plane perpendicular to the optical axis. These flexures follow the same design principles as those used for the spherical mirrors and are illustrated in Fig.~\ref{fig:loft_island}. Multimode fibers are used as the input and output of the optical system thus lateral alignment is sufficient for fiber coupling. Optical components that do not require active degrees of freedom (DOF) are placed in shallow alignment pockets~\footnote{Such pockets can be accurately made using subtractive manufacturing.}. 

Most dimensions of the AMOC incorporate explicit tolerances. Features requiring tighter control were assigned a tolerance of \SI{0.25}{\milli\meter}, while less critical dimensions were given a \SI{0.3}{\milli\meter} tolerance. Although these values are large compared to those achievable with computer numerical control machining, they reflect both the resolution limits of the Renishaw AM 400 system, whose minimum feature size is approximately \SI{35}{\micro\meter} (Sec.~\ref{sec:manufacturing}), and the manufacturing tolerances of the optical components themselves. In practice, these larger tolerances may permit slight motion of the mounted optics, however, such movement can be eliminated by securing the components using standard optical adhesive techniques.
	
The central support structures that hold the stationary elements are part of the monolithic chassis and are therefore built using AM. Shown in Fig.~\ref{fig:Cadchassis}, these structures are designed such that each optical element pocket is supported by its own tapered structure grown from the baseplate. We call these pillars \textit{Loft Islands}, shown in Fig.~\ref{fig:loft_island}. Their geometry is designed to avoid abrupt changes in cross‑sectional areas from the underlying support to the functional surface at the top. Such sudden area transitions are known to challenge the Renishaw AM 400 and can lead to print failures~\cite{TannousRamy2023}. The tapered profile mitigates this issue by gradually increasing the cross‑section as the structure grows upward, ensuring stable layer‑by‑layer fabrication. To reduce the material usage, each \textit{Loft Island} structure is hollowed internally, providing weight reductions that are difficult to achieve using standard manufacturing techniques while maintaining a fully monolithic structure. 

The center of the \textit{Loft Island} includes a small tabletop feature used to secure flat and prism mirrors. To prevent printing failures, this tabletop is supported by a chamfer grown from the island walls at an angle of \SI{35}{\degree} relative to the vertical, as shown in Fig.~\ref{fig:loft_island_inside1}. This angled support ensures that no unsupported overhangs are introduced during fabrication. The central tabletop is not strictly required if the optical elements are bonded using adhesive, the feature can be removed entirely and is not a critical aspect of the Loft Island design.

\subsubsection{Flexure Design}
\label{sec:flexure}
\begin{figure}[htbp]
       \centering

        \begin{subfigure}[b]{0.25\textwidth}
        \centering
            \includegraphics[width=\textwidth]{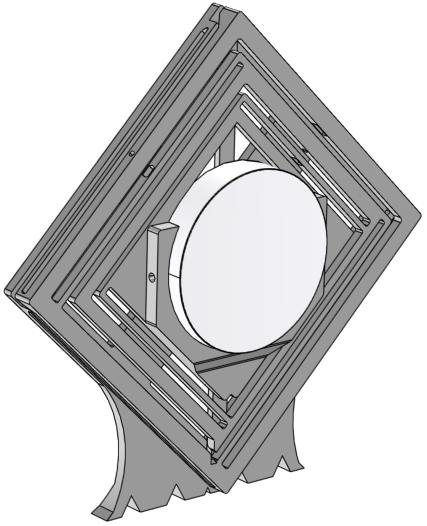}
            \caption{}
        \end{subfigure}
        \begin{subfigure}[b]{0.25\textwidth}
        \centering
            \includegraphics[width=\textwidth]{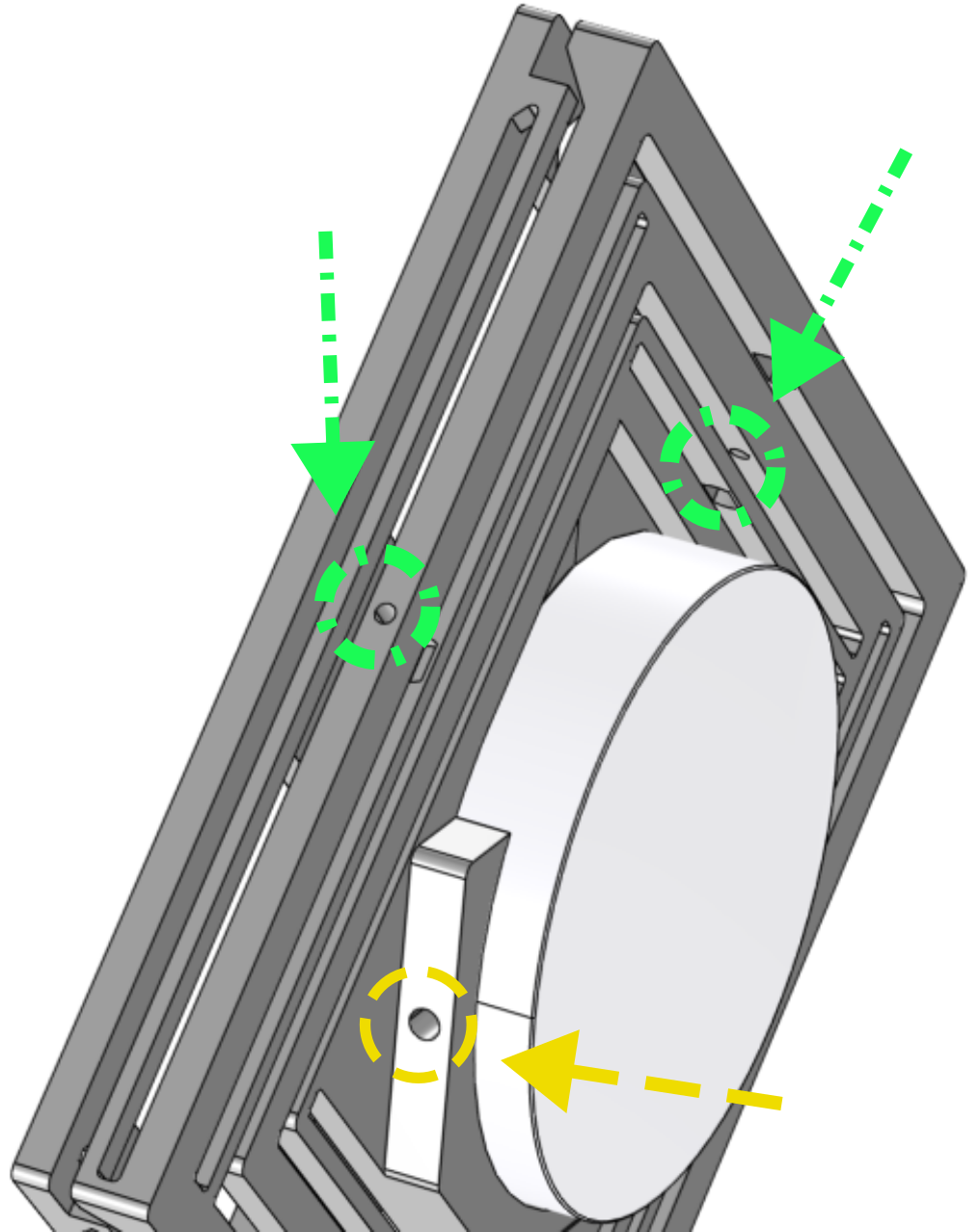}   
            \caption{}
            \label{fig:close_up}
        \end{subfigure}
        \\
        \begin{subfigure}[b]{0.25\textwidth}
            \centering
            \includegraphics[width=\textwidth]{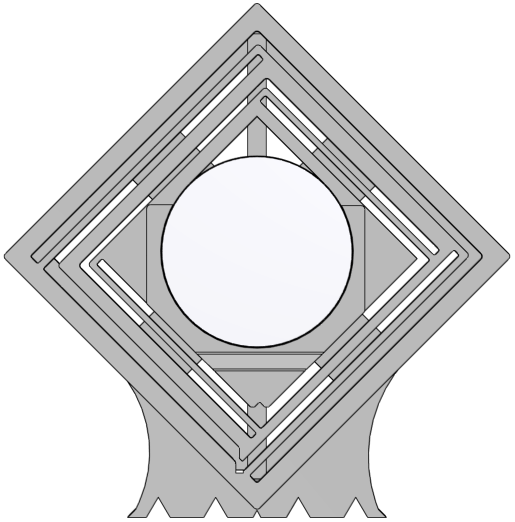}   
            \caption{}
        \end{subfigure}
        \begin{subfigure}[b]{0.25\textwidth}
            \centering
            \includegraphics[width=\textwidth]{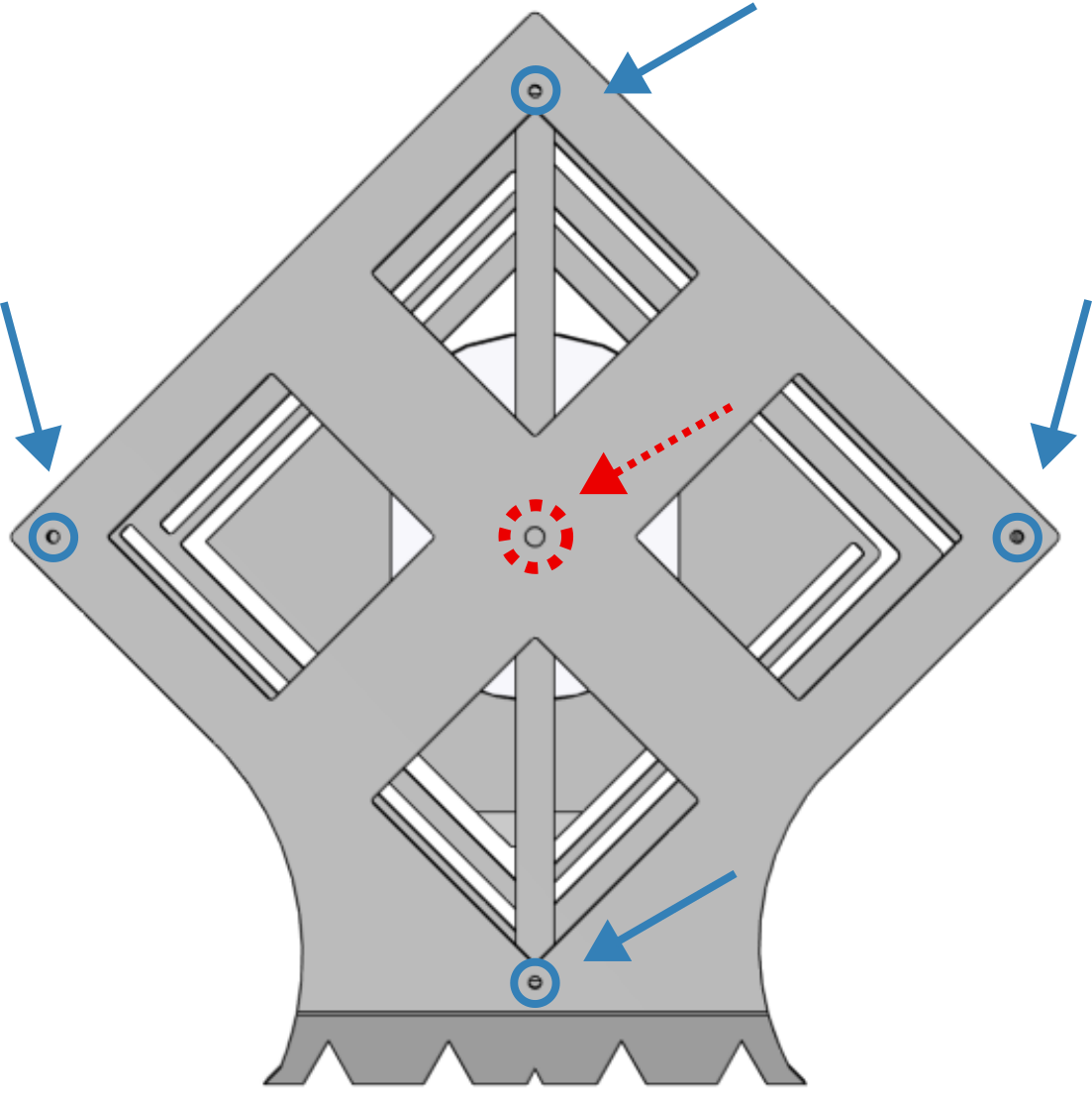}
            \caption{}
            \label{fig:back_flex}
        \end{subfigure}
        
        \caption{Monolithic flexure device for alignment of the folded Offner relay. The spherical mirror rests in the flexure device and set screws push on the structure that is held by the flexure arms. The close up image (\subref{fig:close_up}) shows the single beam connection between back plate and the front of the flexure. Various screw holes are annotated and explained in the main text.}
        \label{fig:flexure}
\end{figure}

The optomechanical flexures are designed to minimize both the physical footprint and material usage of the AMOC. The monolithic flexure used to align the spherical mirrors, shown in Fig.~\ref{fig:flexure}, features a compact geometry capable of providing motion along three orthogonal axes. Its design accounts for the advantages and limitations of additive manufacturing, including the printing tolerances inherent to the process. With these constraints in mind, the flexure is engineered to provide at least \SI{3}{mm} of translation along the spherical mirror’s optical axis and \SI{1.5}{mm} of travel along each of the two perpendicular axes.

COMSOL simulations were used during the design phase to estimate mechanical stress at maximum displacement. Using the yield strength of bulk Invar36 (276~MPa) as a reference, we targeted a minimum safety factor of 1.3 to ensure that the flexure remains within the elastic regime throughout its operational range. In general, flexures can be tailored to provide a desired displacement range, allowing their geometry to be adapted to the specific alignment requirements of the optical system.

The flexures operate by bending the thin beams that connect the center optical mount to the outer frame~\cite{howell_handbook_2013,flexure_iandiorio2023elasto}. Two set screws (annotated in the green dot-dashed markings in Fig.~\ref{fig:close_up}), one for each axis (X-Y), apply force to the central mount, while the compliant beams act as springs to provide controlled motion. A similar beam‑based design is used to achieve translation along the optical axis. In this case, a single folded beam connects the front flexure assembly to a rear plate that is anchored to the baseplate. Axial translation is controlled by five set screws: four located at the corners (blue solid annotation in Fig.~\ref{fig:back_flex}) of the structure and a central “master” screw (red dotted annotation in Fig.~\ref{fig:back_flex}). The master screw provides pure translation along the optical axis, while the corner screws enable tip‑tilt adjustments. The spherical mirror is held within a semi‑open frame and secured using a nylon‑tipped set screw (yellow dashed annotations in Fig.~\ref{fig:close_up}).

\section{Manufacturing Details}
\label{sec:manufacturing}

 As mentioned in Sec.~\ref{sec:additive}, Invar36 is used to manufacture the AMOC. The Invar36 powder used is gas atomized, with a size distribution of \SIrange[]{20}{63}{\micro m} (D10 of \SI{27.7}{\micro\meter}, D50 of \SI{46.4}{\micro\meter}, and D90 of \SI{73.4}{\micro\meter}). The build-plate substrate was stainless steel. To maximize density while minimizing surface roughness and residual stresses, physics-based processing diagrams~\cite{patel2020melting}, X-ray computed tomography, and laser profilometry are used to identify the LPBF processing parameters. For laser powder bed fusion system used in this work (AM400 Renishaw, UK), the beam velocity variable is computed based on two variables, point distance and exposure time. The scan strategy followed this order: scanning of the core using the meander scanning strategy, followed by a single border scan that involves melting of the edge of each contour within a layer specified by the CAD. The important parameters used for manufacturing include a laser power of \SI{250}{\watt}, beam spot diameter of \SI{100}{\micro\meter}, point distance of \SI{60}{\micro\meter}, exposure time of \SI{50}{\micro\second}, a powder layer thickness value of \SI{40}{\micro\meter}, and a hatching distance of \SI{100}{\micro\meter}. The AM 400 system has a fixed minimum spot size at the focal point of \SI{35}{\micro\meter}. To obtain the beam spot diameter of \SI{100}{\micro\meter}, the laser beam focused was moved to a position above the build-plate to create a divergent beam at the laser material interaction plane. Divergent beams help reduce porosity in LPBF parts by reducing the effective beam power density as melting progresses~\cite{patel2021influence}. For border scanning, a laser power of \SI{150}{\watt}, point distance of \SI{40}{\micro\meter}, exposure time of \SI{60}{\micro\second} is used. The build-plate is kept at room temperature during the print, and argon is used as the shielding gas. Table~\ref{tab:manufact} provides a summary of the critical print parameters.

 \begin{table}[htbp]
\centering
\caption{Summary of the relevant manufacturing parameters for the LPBF process.}
\label{tab:manufact}
\begin{tabular}{@{}lc@{}}
\toprule
Parameter               & Value                        \\ \midrule
Powder material                & Invar36 powder               \\
Base plate material         & Stainless steel                       \\
AM system                  & Renishaw AM 400              \\
Laser power             & \SI{250}{\watt}               \\
Beam spot diameter      & \SI{100}{\micro\meter} \\
Exposure time           & \SI{50}{\micro\second}  \\
Powder layer thickness & \SI{40}{\micro\meter}  \\
Hatching distance       & \SI{100}{\micro\meter}  \\
Point distance          & \SI{60}{\micro\meter}  \\ \bottomrule
\end{tabular}
\end{table}

\section{Results}
\label{sec:experiment}
\begin{figure}[htbp]
		\centering
    \begin{subfigure}{0.45\columnwidth}
			\includegraphics[width=\columnwidth]{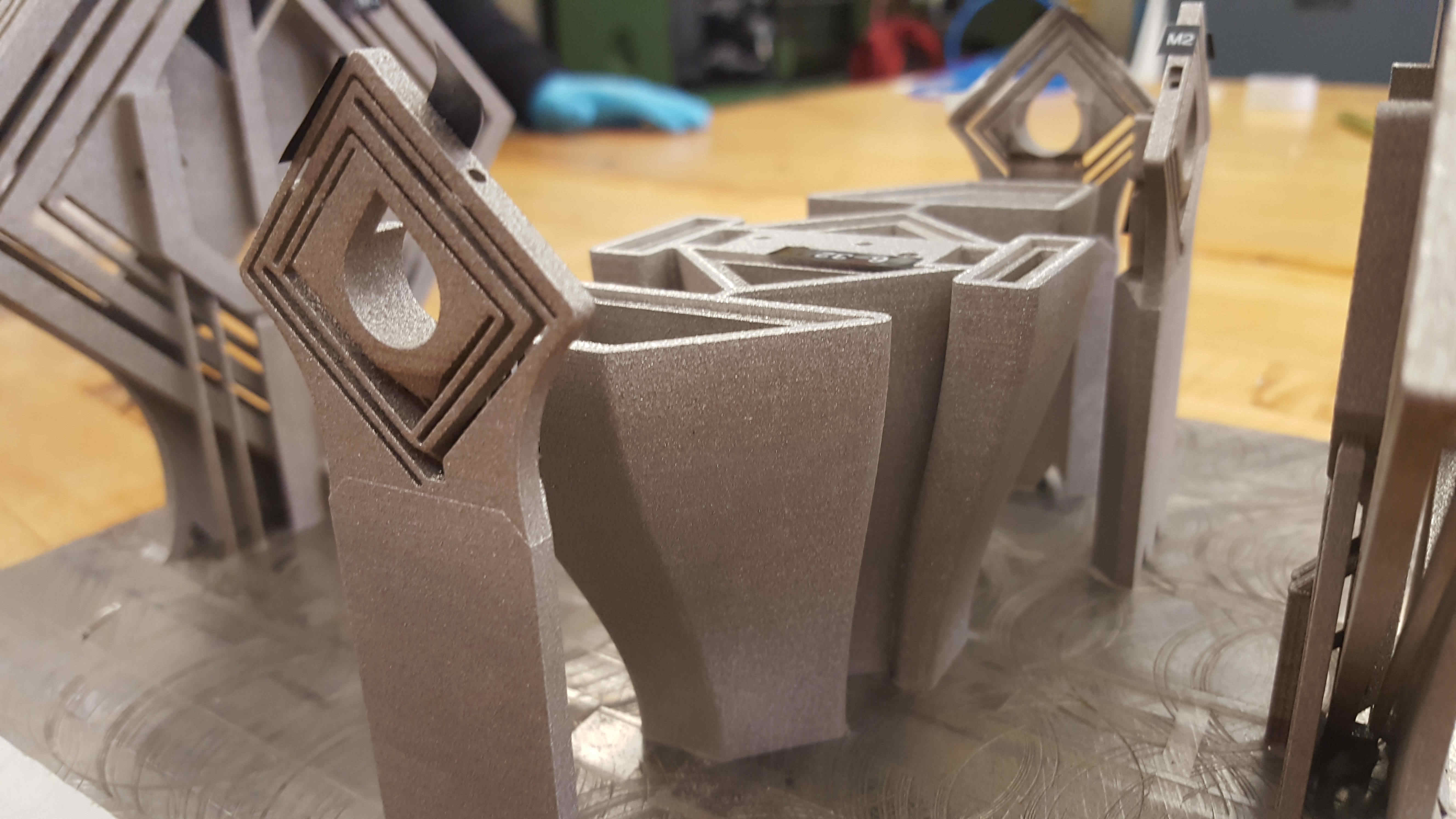}
			\caption{}
			\label{fig:PORI_1}
    \end{subfigure}
    \begin{subfigure}{0.45\columnwidth}
			\centering
			\includegraphics[width=0.85\columnwidth]{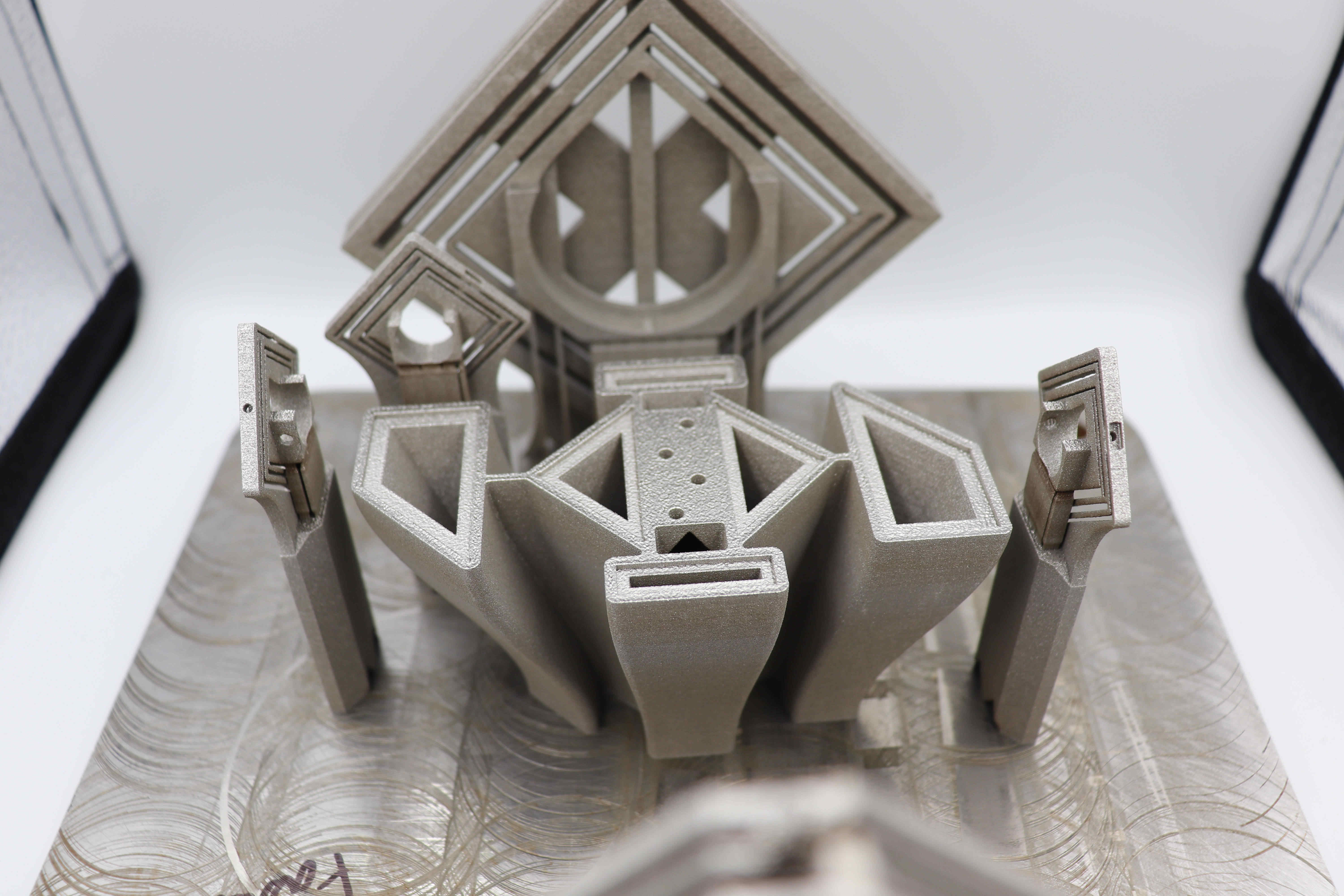}
			\caption{}
			\label{fig:PORI_2}
    \end{subfigure}
    \caption{ (\subref{fig:PORI_1}) and (\subref{fig:PORI_2}) \textit{Loft Island} design printed in combination with the optomechanical flexures.}
    \label{fig:PORI_prints}
\end{figure}

Figure~\ref{fig:PORI_prints} shows the completed AMOC print prior to the installation of any optical components. The chassis is printed as a single monolithic structure, with temporary support features visible in the flexure regions in Fig~\ref{fig:PORI_prints}. The temporary supports are manually removed after fabrication. Additional post-fabrication processes include tapping of the screw holes required for the alignment and securing of optical components. Although the \textit{Loft Island} design substantially increased the print success probability, the final AMOC print was not fully complete, failing during the last few layers due to powder dosing limitations. 
The error responsible for terminating the \SI{20}{\hour} print is minor and not specific to the design, and could be readily addressed in a subsequent fabrication run. However, due to budgeting constraints, machine scheduling, and project timelines, a second print could not be commissioned. As a result of the incomplete print, the larger spherical‑mirror flexures were missing portions of their four‑arm structure, leading to deviations in stiffness and range of motion relative to the simulated design.

Maintaining a stable relative phase between the two arms of an unbalanced interferometer requires minimizing both thermal drift and mechanical excitation. In the AMOC, the dominant sources of mechanical instability arise from the vibration modes of the flexures, therefore any environmental vibrations near their natural frequencies can excite them, resulting in resonant amplification. Additionally, thermal expansion of the chassis likewise perturbs the optical path difference and can degrade interference visibility. For these reasons, the vibrational and thermal behavior of the monolithic structure were evaluated during the design stage using finite‑element analysis (FEA), with results presented in Sec.~\ref{sec:mech} and Sec.~\ref{sec:therm}. The optical performance of the PORI is then validated experimentally by measuring the interference visibility using the optical elements aligned and housed within the AMOC (Sec.~\ref{sec:optical}).

\subsection{Mechanical Stability}
\label{sec:mech}

The mechanical stability of the AMOC is governed primarily by the behavior of its flexure mounts. The flexure range of motion was first evaluated using finite‑element analysis in COMSOL, which predicted that the stresses remain below the yield strength of bulk Invar36 for displacements up to \SI{1}{mm} along each lateral axis and \SI{3}{mm} along the optical axis. Surprisingly, the printed flexures exhibited a substantially larger elastic travel exceeding \SI{5}{mm}, while still operating within the compliant regime. This discrepancy reflects potential differences between the bulk material properties and those of the microstructure produced by the LPBF AM process, suggesting that the effective yield strength may be influenced by print parameters and subsequent heat treatments~\cite{rishmawi2022effects,hitzler2022_INVAR_thermal,HUANG2024104025}. A detailed investigation of these effects is beyond the scope of the present work, and the flexure design is therefore based on the nominal properties of bulk Invar36.

The vibration behavior of the flexures is set by the elastic stiffness of their beams, which determines both the resonance frequencies and the usable range of motion. High stiffness raises the resonant modes and improves mechanical stability but limits travel, whereas low stiffness increases susceptibility to low‑frequency environmental vibrations while enabling larger displacements. In this design, we selected a stiffness that provides sufficient range for alignment while maintaining adequate vibrational robustness for experimental operation. In practice, once alignment is achieved, the flexures can be immobilized with adhesive, however, their short‑ and long‑term stability remains important for preserving alignment during adjustment. These stability characteristics are measured and compared against a conventional optomechanical mount (ThorLabs KM200PM).
    
The short‑term stability of the flexure mounts was evaluated to quantify their response to ambient vibrations and external disturbances. A piezo‑ceramic stack placed between the lateral beams recorded the flexure motion, and the background vibration spectrum was obtained via FFT analysis. Both the AMOC flexure and the reference mount (ThorLabs KM200PM) exhibited electromagnetic noise near 50-60 Hz but the printed flexure showed additional low‑frequency components at $32$, $120$, $176$, and $296$ \si{Hz}, consistent with its lower stiffness. These measured frequency components differed from the COMSOL eigenfrequency analysis, reflecting the increased compliance of the additively manufactured Invar36. 

Damping behavior was quantified by the time required for the oscillation amplitude to decay to $1/e$ following a perturbation. The printed flexure exhibited an average dampening time of \SI{1.55}{\second}, which is unusually long for an optomechanical mount.  However, due to the project timeline, limited budget and availability of the AM 400 machine, it was not possible to fabricate a revised model or to investigate microstructural tailoring of the material via process parameters and heat treatments. Instead, temporary compressible dampers were inserted between the flexure beams. The dampeners removed several of the low frequency components, leaving only the $296$ \si{Hz} mode and reduced the average dampening time to \SI{0.34}{\second}. For comparison, the KM200PM mount showed negligible low frequency content and exhibited an average dampening time of \SI{0.08}{\second}. Further optimization of flexure stiffness will be required to improve the vibrational robustness of future AMOC designs.

    
The long‑term stability of the flexures was evaluated by monitoring the position of a reflected beam over a 12‑hour period while the mounts were held under load. The beam centroid was recorded once per minute using a WinCamD‑LCM profiler, and no composite dampers were used during this test. In the horizontal direction, the 3D‑printed flexure exhibited stability comparable to the reference KM200PM mount, with standard deviations of $0.8677$ and $0.8625$ \unit{\um}, respectively. In contrast, the stability of the flexure in the vertical axis was worse as the standard deviation increased to $0.9912$ \unit{\um}, compared to $0.2413$ \unit{\um} for the control mount. This increased drift is consistent with the lower stiffness of the additively manufactured flexure, which makes it more susceptible to gravitational sag over extended periods.

\subsection{Thermal Stability}
\label{sec:therm}
Although external insulation can help mitigate temperature‑induced phase drift in an interferometer, our focus here is on the intrinsic thermal stability of the chassis material, specifically, its coefficient of thermal expansion. To minimize thermally driven changes in optical‑path length, the chassis is fabricated from Invar36, which offers exceptionally low thermal expansion ($1.3\times10^{-6}$~\si{\per\kelvin} near room temperature \cite{guillaume1904invar,carpenter_invar36}) while maintaining adequate elastic strength for compliant mechanisms. Ideally, both the chassis and the baseplate would be constructed entirely from Invar36 to ensure uniform thermal behavior. However, due to the higher cost of Invar, alternative build‑plate materials were also considered in the thermal analysis.

The thermal behavior of the monolithic chassis was evaluated using steady‑state COMSOL simulations over a \qty{20}{\degreeCelsius} temperature range. The resulting changes in optical‑path length between selected points of interest and the associated mechanical stress were analyzed for three material configurations: Invar36 flexures on an Invar36 base, Invar36 flexures on a stainless‑steel base, and Invar36 flexures on an aluminum‑6061 base. This comparison highlights the impact of base‑material thermal expansion on interferometer path stability and overall structural loading.

	
The thermal stability is quantified by measuring the change in the distance the light ray travels for each path of the interferometer ($\Delta{L}$) compared to design parameters. A high thermal stability would result in a small $\Delta{L}$. The path deviations are calculated assuming a horizontal aligned beam, i.e. no vertical components of thermal expansion are considered. We also specifically report the change in distance between the two large flexures housing the spherical mirrors
as thermal expansion in this axis is more detrimental to the system. In addition, stress builds up at assembly points at the intersection of baseplate and flexures, and the overall out-of-plane deformations are also investigated. 

\begin{table}[htbp]
	\centering
	\caption{Summary of the thermal expansion simulations of the PORI for an Invar, stainless steel (SS), and aluminum base plate. The last two columns show the resulting phase shift caused by the expansion as the number of wavelengths at \SI{800}{\nano\meter} and at \SI{1550}{\nano\meter}.}
	\label{tab:therm_sim_res}
        
	\begin{tabularx}{0.8\columnwidth}{L{0.5}C{0.5}C{0.5}C{0.5}C{0.5}C{0.5}}
		\toprule
		Metal    & $\Delta$ T [\si{\degreeCelsius}]  & $\Delta$ L [\si{\micro\meter}] & Delay change [\si{\femto\second}] &  $\lambda$ @ 800\si{\nano\meter} & $\lambda$ @ 1500\si{\nano\meter} \\ \midrule
		Invar    & 5  & 3.19       & 1.27                  & 4.8        & 2.5         \\
		Invar    & 20 & 12.53      & 4.18                  & 15.6       & 8.3         \\
		SS       & 5  & 29.4       & 9.79                  & 36.7       & 19.6        \\
		SS       & 20 & 95.9       & 31.95                 & 119.8      & 63.9        \\
		Aluminum & 5  & 93.8       & 31.26                 & 117.2      & 62.5        \\
		Aluminum & 20 & 303.5      & 101.18                & 379.4      & 202.4       \\ \bottomrule
	\end{tabularx}
\end{table}

Table.~\ref{tab:therm_sim_res} shows unsurprisingly that the Invar36 base plate performs the best compared to the stainless steel and aluminum base plates. The $\Delta{L}$ of the Invar36 base plate is approximately $\Delta{L_5}=$~3.19 \si{\micro\meter} and $\Delta{L_{20}}=$~12.53 \si{\micro\meter} for \qty{5}{\degreeCelsius} and \qty{20}{\degreeCelsius} change in overall temperature of the chassis, respectively. Converting to the time (phase) domain, this deviation from the design parameters correspond to $\Delta{T_5}=$~1.27~\si{fs} ($4.8\lambda$ @800nm, $2.5\lambda$ @1500nm) and $\Delta{T_{20}}=$~4.18~\si{fs} ($15.6\lambda$ @800nm, $8.3\lambda$ @1500nm) increase in relative path delay between the two interferometer arms, respectively. The maximum overall change in distance between the two spherical mirrors is calculated to be approximately 1.3 micrometers. No out-of-plane deformations are observed, and mechanical stress build up is present but trivial due to matching thermal expansion coefficients of the base plate and flexures. The mechanical stress at the base plate-flexure joints is in the millipascal range even for \qty{20}{\degreeCelsius} temperature change, which is small compared to the stainless steel and aluminum base plates that exhibit megapascal ranges of stress.

For the stainless steel (SS) base plate, the $\delta{L}$ of the optical path length was approximately $\Delta{L_{5}}=$~29.4~\unit{\um} and $\Delta{L_{20}}=$~95.9~\unit{\um}, respectively. Corresponds to $\Delta{T_{5}}=$~$9.79$~ \si{fs} ($36.7\lambda$ @800nm, $19.6\lambda$ @1500nm) and $\Delta{T_{20}}=$~$31.95$~ \si{fs} ($119.8\lambda$ @800nm, $63.9\lambda$ @1500nm) increase in relative delay between the two interferometer paths. The change in distance between the two large flexures is 17.2 \unit{\um} for even the \qty{5}{\degreeCelsius} temperature difference, significantly higher than the Invar36. Out-of-plane deformations are also significant, causing $0.80$ \unit{\um} relative warping between the two central beam displacers and approximately 0.63 \unit{\um} warping for the large flexure mounts compared to the initial base plate plane, for a \qty{5}{\degreeCelsius} temperature change.

For the aluminum (6061 alloy) base plate, the $\Delta{L}$ of the interferometer is calculated to be approximately 93.8 and 303.5 \unit{\um} for \qty{5}{\degreeCelsius} and \qty{20}{\degreeCelsius} temperature change, respectively. The deviation from the design parameters corresponds to 31.26 ($117.2\lambda$ @800nm, $62.5\lambda$ @1500nm) and 101.18 ($379.4\lambda$ @800nm, $202.4\lambda$ @1500nm) \si{fs} increase in the relative path delay between for \qty{5}{\degreeCelsius} and \qty{20}{\degreeCelsius} temperature change, respectively. The maximum delta for large flexure distance is $31.5$ \unit{\um}. Out-of-plane deformations are 1.67 \unit{\um} relative warping between the beam displacers and approximately $0.93$ \unit{\um} warping for the large flexures compared to initial base plate plane, for a \qty{5}{\degreeCelsius} temperature change. Evidently, the path length deviations for stainless steel and aluminum base plate setups are too large to preserve phase stability of the interferometer, while warping would cause the beam to go out of alignment. 

As expected from the warping behaviour, stress build up at the attachment boundary layer between the SS (aluminum) base plate and Invar36 flexures were high, reaching approximately \SI{98.8}{\mega\pascal} (148.7 MPa) at the attachment points of the large flexures for only 5 \si{\degreeCelsius} temperature change. At a larger temperature change of 17 \unit{\degreeCelsius} (12 \unit{\degreeCelsius}), stress build up reaches the yield strength of Invar36 $\approx276$ MPa. Considering the imperfect joint between the base plate and the AM structures, the chassis would be expected to fail before ambient temperature change reaches this magnitude. Overall, the thermal simulations demonstrate that a system made of Invar will have the best thermal stability.

\subsection{Optical Performance}
\label{sec:optical}
\begin{figure}[htbp]
    \centering
    \includegraphics[width=0.65\textwidth]{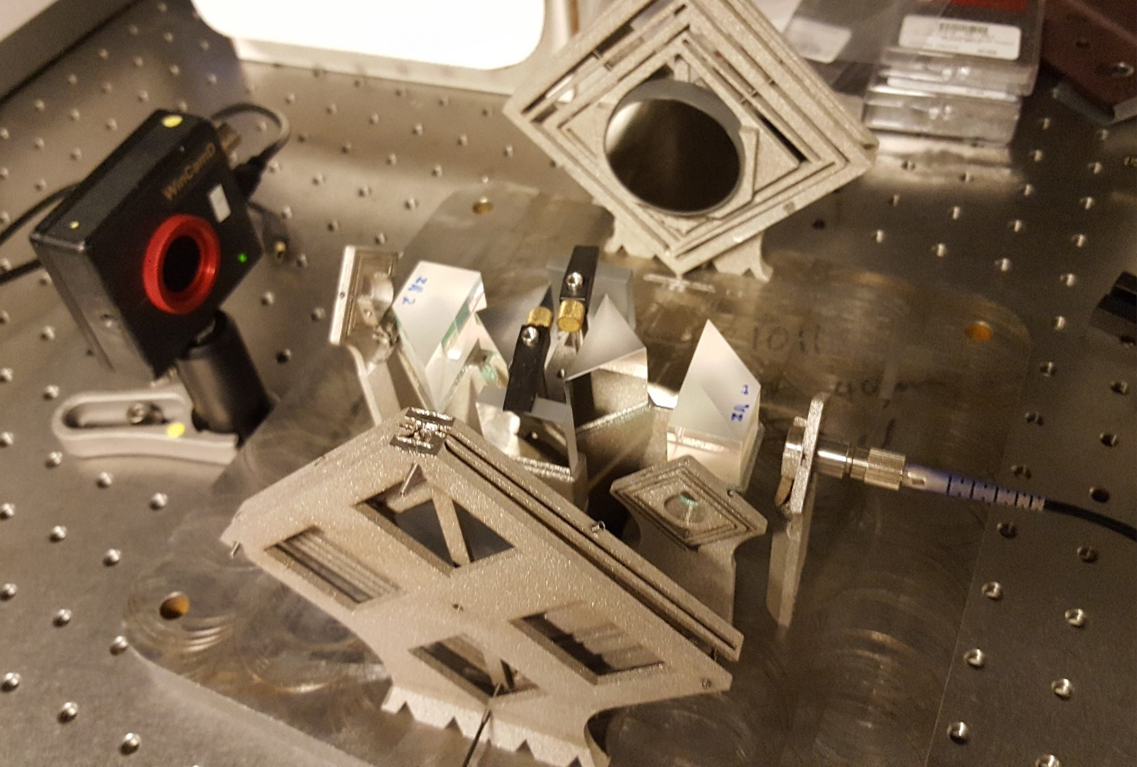}
    \caption{The PORI with the Offner relay interferometer optics in place. The flexures and optical supports are constructed using Invar36 while the baseplate is composed of stainless steel.}
    \label{fig:chassis}
\end{figure}

\begin{figure}[htbp]
    \centering
    \begin{subfigure}[b]{0.495\textwidth}
        \centering
            \includegraphics[width=0.99\textwidth]{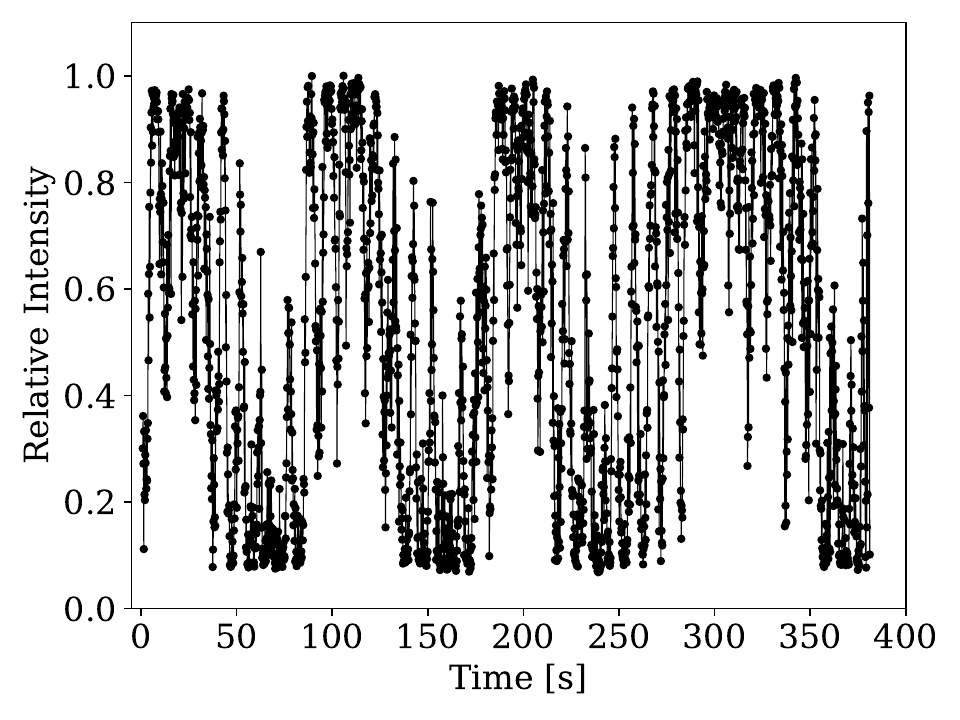}   
            \caption{Single-mode signal}
            \label{fig:smf-vis}
    \end{subfigure}
    \begin{subfigure}[b]{0.495\textwidth}
        \centering
            \includegraphics[width=0.99\textwidth]{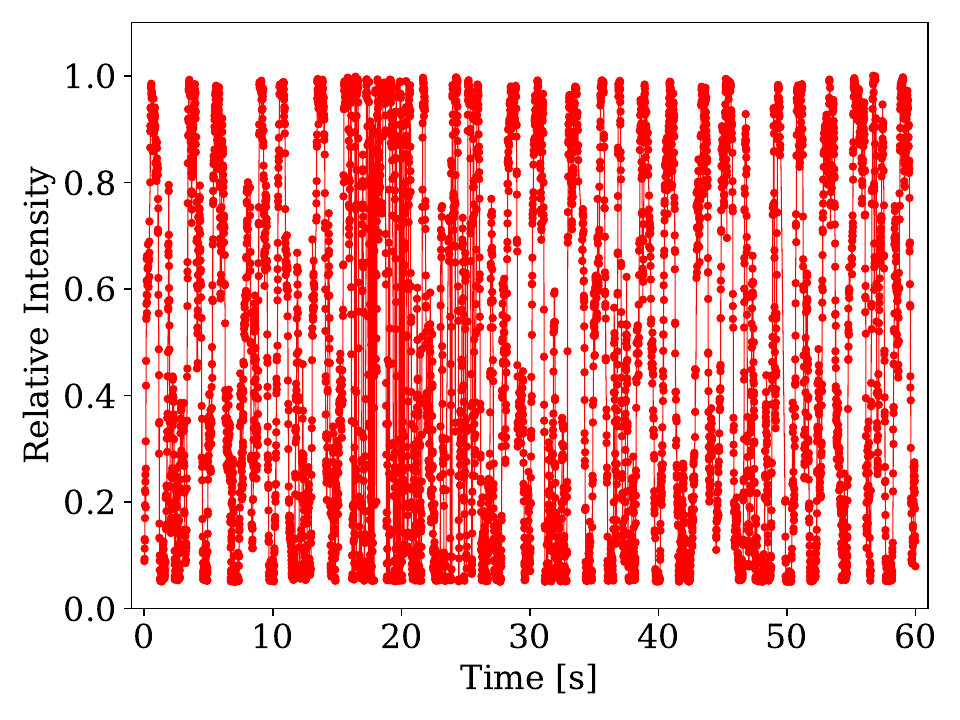}   
            \caption{multimode signal}
            \label{fig:mmf-vis}
    \end{subfigure}
    \caption{The relative intensity of the PORI for both the single-mode case (a) and the multimode case (b). The relative intensity values are normalized to the maximal value of the data set. The zero intensity level is the background level of the camera.}
    \label{fig:vis}    
\end{figure}

The fully assembled monolithic chassis, populated with the optical elements of the Offner‑relay interferometer, is shown in Fig.~\ref{fig:chassis}, forming a complete PORI system. To our knowledge, this is the first monolithic interferometer chassis that includes optomechanical flexures that allow for the alignment of the optical system. The optical performance of the system is quantified using interference visibility. The interference visibility is defined as,
    \begin{equation}
        \label{eq:vis}
        V=\frac{I_{\textit{max}}-I_{\textit{min}}}{I_{\textit{max}}+I_{\textit{min}}},
    \end{equation}
where $I_{\textit{max(min)}}$ are the maximum (minimum) measured intensity. The system is tested with two types of signals, single-mode and multimode laser light, though multimode light is more representative of a quantum sensing scenario, the intended use of the interferometer~\cite{sajeed_observing_2021}. A continuous wave \SI{785}{nm} laser is passed through a single-mode or multimode fiber before passing through the PORI. A piezo-ceramic actuator is used to vary the path length of one of the PORI paths to observe interference fringes. However, in Fig.~\ref{fig:smf-vis} and Fig.~\ref{fig:mmf-vis}, the results are difficult to interpret due to noisy and non-periodic fluctuations caused by the vibrational instabilities of the flexure mounts which can cause the relative phase between the two paths of the PORI to rapidly fluctuated during the measurement. Nonetheless, the measured maximum visibility for both the single-mode and multimode signals are $0.861\pm 0.025$ and $0.878\pm 0.038$ respectively. The lower visibility performance compared to Ref.~\cite{tannous2026all} can be attributed to the flexure instabilities caused by the print failure that made it difficult to properly align the system. Nonetheless, the visibility with the multimode signals is much greater than with the results obtained with a standard Michelson interferometer ($0.536 \pm 0.023$) using the same spot size (\SI{0.5}{\milli\meter}) and experimental conditions as the PORI.

\section{Discussion and Conclusion}
\label{sec:discussion}

In this work we presented a monolithic chassis for a field-widened interferometer that is constructed entirely using additive manufacturing. The monolithic design included optomechanical flexure components for alignment of the interferometer and is printed using low thermal expansion materials, increasing the thermal stability of the device. The prototyping study had four essential design criteria that the optomechanical structure should meet. (1) A reduced physical form-factor: custom, low‑profile optomechanical components were developed to limit the footprint increase introduced by the monolithic chassis relative to the original optical layout. Excluding the height contribution of the PORI, the complete system—constructed with standard 2‑inch optics—exhibited a $13\%$ reduction in physical footprint compared to off-the-self optomechanical components. Further reductions to the footprint are achievable by decreasing the size of the optical components; for example, using 1‑inch optics reduces the height of the flexure stages such that the physical volume of the device is 20\% smaller than a comparable bench-top system. Additional improvements are possible through custom optical elements and further refinement of the flexure designs. (2) A monolithic assembly-free structure: all optomechanical components are manufactured into a monolithic structure with minimal additional post manufacturing processes are required to assemble the monolithic chassis beyond the removal of temporary supports and the addition of the required screws. (3) Adjustable mounts for alignment: optomechanical flexures are included in the monolithic structure and were capable of aligning the system to achieve suitable optical performance. (4) Vibrational and thermal stability: the vibrational stability of the device could use some improvements as the flexures exhibited frequency eigenmodes in the low frequency regime. Although some addition composite dampening material is used to reduce vibrational response of the flexures, future work would be to improve the vibrational stability of the device through improved flexure designs and material tailoring. Greater vibrational stability should also lead to improved long-term stability of the flexures. For the thermal stability, using a monolithic design with a relatively uniform material reduced the complexity of the thermal FEA study. Furthermore, the material selection of Invar36 greatly reduces the temperature stability requirements of any insulation or temperature stabilization systems that would be used in combination with the interferometer to maintain the a stable relative phase. Future studies would be to further investigate the performance of the PORI under representative environmental conditions.

A major future area of improvement is to reduce the size of the AMOC by using smaller custom optical components. The AMOC in this exploratory study is designed to use standard off-the-shelf optical components of 1 inch and 2 inch diameters. Designing for such large optical components is not the optimal approach and likely contributed to the print failure as a large chassis was required. Using more compact or custom optics would ultimately reduce the required size and height of the structure that could increase the success of the print. Furthermore, smaller optics would reduce the required size of the flexure mounts which in turn would improve the vibrational stability of the system. Regardless, any reduction in the physical footprint of the device is helpful for applications that are resource limited.

Budgetary and project‑timeline constraints prevented us from pursuing additional fabrication runs beyond the initial simulations and design work~\cite{TannousRamy2023}. Nonetheless, the resulting chassis was largely functional, though the missing upper layers introduced measurable reductions in mechanical and optical stability, as discussed in Secs.~\ref{sec:mech} and~\ref{sec:optical}. Continued improvements in additive‑manufacturing accuracy (surface roughness, distortion compensation, beam compensation for geometric fidelity of fine features) will allow future designs to reduce the flexure travel range and rely more heavily on precise, repeatable AM geometries.

Despite these limitations, the \textit{Loft Island} chassis demonstrates a compelling approach to fabricating optical assemblies. In particular, the direct integration of flexures into a monolithic structure—something exceedingly difficult to achieve with conventional machining—highlights the unique capabilities of AM for optomechanical design. We anticipate that this work will open new pathways for merging additive manufacturing with optical and quantum technologies. A long‑term goal is the direct incorporation of optical surfaces and components into the printed structure itself. Achieving such fully integrated systems with a single manufacturing method will require substantial further study, but hybrid approaches that combine AM with conventional machining offer a practical near‑term route toward this vision.

\begin{acknowledgments}
	We graciously thank Dr. Shihan Sajeed and Dr. Youn Seok Lee for helpful discussion. RT would like to thank the National Science and Engineering Research Council of Canada, Canadian Graduate Scholarship for personal funding. This project was funded by the Deference Research and Development Canada IDEAS program.
\end{acknowledgments}

\bibliography{references}

@article{salazar20173dtoolbox,
  title={A 3D printed toolbox for opto-mechanical components},
  author={Salazar-Serrano, Luis Jos{\'e} and P. Torres, Juan and Valencia, Alejandra},
  journal={PloS one},
  volume={12},
  number={1},
  pages={e0169832},
  year={2017},
  publisher={Public Library of Science San Francisco, CA USA}
}

@article{winters20183dCageSystem,
  title={3D printable optomechanical cage system with enclosure},
  author={Winters, Brandon J and Shepler, David},
  journal={HardwareX},
  volume={3},
  pages={62--81},
  year={2018},
  publisher={Elsevier}
}

@article{zhang2013openSource,
  title={Open-source 3D-printable optics equipment},
  author={Zhang, Chenlong and Anzalone, Nicholas C and Faria, Rodrigo P and Pearce, Joshua M},
  journal={PloS one},
  volume={8},
  number={3},
  pages={e59840},
  year={2013},
  publisher={Public Library of Science San Francisco, USA}
}

@article{hussain2018designPhone,
  title={Design of a 3D printed compact interferometric system and required phone application for small angular measurements},
  author={Hussain, I and Nath, P},
  journal={Review of Scientific Instruments},
  volume={89},
  number={10},
  pages={103111},
  year={2018},
  publisher={AIP Publishing LLC}
}

@inproceedings{kranert20203dLowCost,
  title={3D-printed, low-cost, lightweight optomechanics for a compact, low-power solid-state amplifier system},
  author={Kranert, Fabian and Budde, Jana and Neef, Philipp and Bernhard, Robert and Lammers, Marius and Rettschlag, Katharina and Grabe, Tobias and Wienke, Andreas and Neumann, J{\"o}rg and Wiche, Henning and others},
  booktitle={Components and Packaging for Laser Systems VI},
  volume={11261},
  pages={1126105},
  year={2020},
  organization={International Society for Optics and Photonics}
}

@inproceedings{mammini20153dOpticalMount,
  title={3D-additive manufactured optical mount},
  author={Mammini, Paul V and Ciscel, David and Wooten, John},
  booktitle={Optomechanical Engineering 2015},
  volume={9573},
  pages={957309},
  year={2015},
  organization={International Society for Optics and Photonics}
}

@inproceedings{kranert2021Quasi,
  title={Quasi-monolithic laser system based on 3D-printed optomechanics},
  author={Kranert, Fabian and Budde, Jana and Hinkelmann, Moritz and Wienke, Andreas and Neumann, J{\"o}rg and Kracht, Dietmar and Lachmayer, Roland},
  booktitle={Components and Packaging for Laser Systems VII},
  volume={11667},
  pages={116670L},
  year={2021},
  organization={International Society for Optics and Photonics}
}

@article{sajeed_observing_2021,
	title = {Observing quantum coherence from photons scattered in free-space},
	urldate = {2021-03-15},
	journal = {arXiv:2103.00298 [quant-ph]},
	author = {Sajeed, Shihan and Jennewein, Thomas},
	month = feb,
	year = {2021},
}

@article{jin2018demonstration,
  title={Demonstration of analyzers for multimode photonic time-bin qubits},
  author={Jin, Jeongwan and Agne, Sascha and Bourgoin, Jean-Philippe and Zhang, Yanbao and L{\"u}tkenhaus, Norbert and Jennewein, Thomas},
  journal={Physical Review A},
  volume={97},
  number={4},
  pages={043847},
  year={2018},
  publisher={APS}
}

@article{madkhaly2021performance,
  title={Performance-optimized components for quantum technologies via additive manufacturing},
  author={Madkhaly, SH and Coles, LA and Morley, C and Colquhoun, CD and Fromhold, TM and Cooper, N and Hackerm{\"u}ller, L},
  journal={arXiv preprint arXiv:2102.11874},
  year={2021}
}

@article{cohen2019thresholded,
  title={Thresholded quantum LIDAR: exploiting photon-number-resolving detection},
  author={Cohen, Lior and Matekole, Elisha S and Sher, Yoni and Istrati, Daniel and Eisenberg, Hagai S and Dowling, Jonathan P},
  journal={Physical Review Letters},
  volume={123},
  number={20},
  pages={203601},
  year={2019},
  publisher={APS}
}

@inproceedings{haase202132x24,
  title={A 32x24 pixel SPAD detector system for LiDAR and quantum imaging},
  author={Haase, Jan F and Grollius, Sara and Grosse, Simon and Buchner, Andre and Ligges, Manuel},
  booktitle={Photonic Instrumentation Engineering VIII},
  volume={11693},
  pages={116930M},
  year={2021},
  organization={International Society for Optics and Photonics}
}

@inproceedings{liu2020inspiring,
  title={Inspiring radar from quantum-enhanced LiDAR},
  author={Liu, Han and Helmy, Amr and Balaji, Bhashyam},
  booktitle={2020 IEEE International Radar Conference (RADAR)},
  pages={964--968},
  year={2020},
  organization={IEEE}
}

@article{mcgregor2021architectural,
  title={Architectural bone parameters and the relationship to titanium lattice design for powder bed fusion additive manufacturing},
  author={McGregor, Martine and Patel, Sagar and McLachlin, Stewart and Vlasea, Mihaela},
  journal={Additive Manufacturing},
  volume={47},
  pages={102273},
  year={2021},
  publisher={Elsevier}
}

@article{brock2021relative,
  title={Relative performance of additively manufactured and cast aluminum alloys},
  author={Brock, Lisa and Ogunsanya, Ibrahim and Asgari, Hamed and Patel, Sagar and Vlasea, Mihaela},
  journal={Journal of Materials Engineering and Performance},
  volume={30},
  number={1},
  pages={760--782},
  year={2021},
  publisher={Springer}
}

@article{DebRoy2019Scientific,
	journal={Nature Materials},
	doi={10.1038/s41563-019-0408-2},
	issn={1476-4660},
	number=10,
	note={number: 10
publisher: Nature Publishing Group},
	title={Scientific, technological and economic issues in metal printing and their solutions},
	volume=18,
	author={DebRoy, T. and Mukherjee, T. and Milewski, J. O. and Elmer, J. W. and Ribic, B. and Blecher, J. J. and Zhang, W.},
	pages={1026--1032},
	date={2019-10},
	year=2019,
	month=10,
}

@misc{3DPrinti82:online,
author = {General Electric},
title = {3D Printing in Aerospace \& Aviation | GE Additive},
howpublished = {\url{https://www.ge.com/additive/additive-manufacturing/industries/aviation-aerospace}},
year = {2021},
note = {(Accessed on 06/13/2021)}
}

@misc{Worldfir36:online,
author = {CSIRO},
title = {World-first surgery saves cancer patient's leg},
howpublished = {\url{https://www.csiro.au/en/research/health-medical/biomedical/titanium-heel}},
month = {1},
year = {2021},
note = {(Accessed on 06/13/2021)}
}

@misc{Fraunhof28:online,
author = {Whittaker, Paul},
title = {Fraunhofer ILT and Concept Laser develop Additive Manufacturing for the automotive industry},
howpublished = {\url{https://www.pm-review.com/fraunhofer-ilt-and-concept-laser-develop-additive-manufacturing-for-the-automotive-industry/}},
month = {11},
year = {2012},
note = {(Accessed on 06/13/2021)}
}

@article{khorasani2020review,
  title={A review of technological improvements in laser-based powder bed fusion of metal printers},
  author={Khorasani, AmirMahyar and Gibson, Ian and Veetil, Jithin Kozhuthala and Ghasemi, Amir Hossein},
  journal={The International Journal of Advanced Manufacturing Technology},
  volume={108},
  number={1},
  pages={191--209},
  year={2020},
  publisher={Springer}
}

@article{patel2020melting,
  title={Melting modes in laser powder bed fusion},
  author={Patel, Sagar and Vlasea, Mihaela},
  journal={Materialia},
  volume={9},
  pages={100591},
  year={2020},
  publisher={Elsevier}
}

@book{howell_handbook_2013,
	address = {Chichester, West Sussex, United Kingdom ; Hoboken},
	title = {Handbook of compliant mechanisms},
	isbn = {978-1-119-95345-6},
	publisher = {John Wiley \& Sons, Inc},
	editor = {Howell, Larry L. and Magleby, Spencer P. and Olsen, Brian M.},
	year = {2013}
}

@article{sing2016laser,
  title={Laser and electron-beam powder-bed additive manufacturing of metallic implants: A review on processes, materials and designs},
  author={Sing, Swee Leong and An, Jia and Yeong, Wai Yee and Wiria, Florencia Edith},
  journal={Journal of Orthopaedic Research},
  volume={34},
  number={3},
  pages={369--385},
  year={2016},
  publisher={Wiley Online Library}
}

@article{thompson2016design,
  title={Design for Additive Manufacturing: Trends, opportunities, considerations, and constraints},
  author={Thompson, Mary Kathryn and Moroni, Giovanni and Vaneker, Tom and Fadel, Georges and Campbell, R Ian and Gibson, Ian and Bernard, Alain and Schulz, Joachim and Graf, Patricia and Ahuja, Bhrigu and others},
  journal={CIRP annals},
  volume={65},
  number={2},
  pages={737--760},
  year={2016},
  publisher={Elsevier}
}

@article{wang2018additively,
  title={Additively manufactured hierarchical stainless steels with high strength and ductility},
  author={Wang, Y Morris and Voisin, Thomas and McKeown, Joseph T and Ye, Jianchao and Calta, Nicholas P and Li, Zan and Zeng, Zhi and Zhang, Yin and Chen, Wen and Roehling, Tien Tran and others},
  journal={Nature materials},
  volume={17},
  number={1},
  pages={63--71},
  year={2018},
  publisher={Nature Publishing Group}
}

@book{saito1978physics,
  title={Physics and applications of invar alloys},
  author={Saito, Hideo},
  year={1978},
  publisher={Maruzen}
}

@article{asgari2018thermal,
  title={On thermal expansion behavior of invar alloy fabricated by modulated laser powder bed fusion},
  author={Asgari, Hamed and Salarian, Mehrnaz and Ma, Henry and Olubamiji, Adeola and Vlasea, Mihaela},
  journal={Materials \& Design},
  volume={160},
  pages={895--905},
  year={2018},
  publisher={Elsevier}
}

@article{patel2021influence,
  title={The influence of divergent laser beams on the laser powder bed fusion of a high reflectivity aluminium alloy},
  author={Patel, Sagar and Chen, Haoxiu and Vlasea, Mihaela and Zou, Yu},
  journal={arXiv preprint arXiv:2105.07920},
  year={2021}
}

@phdthesis{TannousRamy2023,
author={{Tannous, Ramy}},
title={Advancing the robustness of polarization and time bin quantum key distribution for free-space channels},
year={2023},
publisher="UWSpace",
school={University of Waterloo},
url={http://hdl.handle.net/10012/19444}
}

@article{flexure_iandiorio2023elasto,
  title={Elasto-kinematics and instantaneous invariants of compliant mechanisms based on flexure hinges},
  author={Iandiorio, Christian and Salvini, Pietro},
  journal={Micromachines},
  volume={14},
  number={4},
  pages={783},
  year={2023},
  publisher={MDPI}
}

@article{beaman2020additive,
  title={Additive manufacturing review: early past to current practice},
  author={Beaman, JJ and Bourell, David L and Seepersad, CC and Kovar, DJJOMS},
  journal={Journal of Manufacturing Science and Engineering},
  volume={142},
  number={11},
  pages={110812},
  year={2020},
  publisher={American Society of Mechanical Engineers}
}

@article{tannous2026all,
  title={All Reflective Field-widened Unbalanced Interferometer for Quantum Sensing and Communication Applications},
  author={Tannous, Ramy and Sinar, Dogan and Arulpragasam, Tabitha D and Jennewein, Thomas},
  journal={arXiv preprint arXiv:2606.18358},
  year={2026}
}

@article{tannous2025towards,
  title={Towards fully passive time-bin quantum key distribution over moving free-space channels},
  author={Tannous, Ramy and Wu, Wilson and Vinet, St{\'e}phane and Perumangatt, Chithrabhanu and Sinar, Dogan and Ling, Alexander and Jennewein, Thomas},
  journal={Optics Express},
  volume={33},
  number={17},
  pages={35635--35648},
  year={2025},
  publisher={Optica Publishing Group}
}

@article{vinet2026time,
  title={Time-resolved certification of frequency-bin entanglement over multi-mode channels},
  author={Vinet, St{\'e}phane and Clementi, Marco and Bacchi, Marcello and Zhang, Yujie and Giacomin, Massimo and Neal, Luke and Villoresi, Paolo and Galli, Matteo and Bajoni, Daniele and Jennewein, Thomas},
  journal={npj Quantum Information},
  volume={12},
  number={1},
  pages={38},
  year={2026},
  publisher={Nature Publishing Group UK London}
}

@article{bedington2017progress,
  title={Progress in satellite quantum key distribution},
  author={Bedington, Robert and Arrazola, Juan Miguel and Ling, Alexander},
  journal={npj Quantum Information},
  volume={3},
  number={1},
  pages={30},
  year={2017},
  publisher={Nature Publishing Group UK London}
}

@article{tian2024experimental,
  title={Experimental demonstration of drone-based quantum key distribution},
  author={Tian, Xiao-Hui and Yang, Ran and Liu, Hua-Ying and Fan, Pengfei and Zhang, Ji-Ning and Gu, Changsheng and Chen, Mengwen and Hu, Mingzhe and Lu, Feng-Yu and Zhu, Chunxi and others},
  journal={Physical Review Letters},
  volume={133},
  number={20},
  pages={200801},
  year={2024},
  publisher={APS}
}

@article{chun2020review_flexure_withAdditive,
  title={A review: additive manufacturing of flexure mechanism for nanopositioning system},
  author={Chun, Heebum and Guo, Xiangyu and Kim, Jung Sub and Lee, ChaBum},
  journal={The International Journal of Advanced Manufacturing Technology},
  volume={110},
  number={3},
  pages={681--703},
  year={2020},
  publisher={Springer}
}

@article{DUPLESSIS2022100918_metal_AM_devices,
title = {Properties and applications of additively manufactured metallic cellular materials: A review},
journal = {Progress in Materials Science},
volume = {125},
pages = {100918},
year = {2022},
issn = {0079-6425},
doi = {https://doi.org/10.1016/j.pmatsci.2021.100918},
url = {https://www.sciencedirect.com/science/article/pii/S0079642521001420},
author = {Anton {du Plessis} and Nima Razavi and Matteo Benedetti and Simone Murchio and Martin Leary and Marcus Watson and Dhruv Bhate and Filippo Berto}
}

@article{COOPER2021101898_vaccuum,
title = {Additively manufactured ultra-high vacuum chamber for portable quantum technologies},
journal = {Additive Manufacturing},
volume = {40},
pages = {101898},
year = {2021},
issn = {2214-8604},
doi = {https://doi.org/10.1016/j.addma.2021.101898},
url = {https://www.sciencedirect.com/science/article/pii/S2214860421000634},
author = {N. Cooper and L.A. Coles and S. Everton and I. Maskery and R.P. Campion and S. Madkhaly and C. Morley and J. O’Shea and W. Evans and R. Saint and P. Krüger and F. Oručević and C. Tuck and R.D. Wildman and T.M. Fromhold and L. Hackermüller}
}

@article{hitzler2022_INVAR_thermal,
  title={Thermal expansion and temperature-dependent Young’s modulus of Invar fabricated via laser powder-bed fusion},
  author={Hitzler, Leonhard and Shahul Hameed, Muhammed Zubair and Kah, Adrian and Merkel, Markus and Werner, Ewald},
  journal={Progress in Additive Manufacturing},
  volume={7},
  number={3},
  pages={463--470},
  year={2022},
  publisher={Springer}
}

@article{HUANG2024104025,
title = {Anisotropy of microstructure, mechanical properties and thermal expansion in Invar 36 alloy fabricated via laser powder bed fusion},
journal = {Additive Manufacturing},
volume = {82},
pages = {104025},
year = {2024},
issn = {2214-8604},
doi = {https://doi.org/10.1016/j.addma.2024.104025},
url = {https://www.sciencedirect.com/science/article/pii/S221486042400071X},
author = {Guoliang Huang and Gongming He and Ying Liu and Ke Huang}
}

@article{rishmawi2022effects,
  title={The effects of heat treatment on tensile and thermal expansion behavior of laser powder-bed fusion Invar36},
  author={Rishmawi, Issa and Rogalsky, Allan and Vlasea, Mihaela and Salarian, Mehrnaz and Bakhshivash, Soheil},
  journal={Journal of Materials Engineering and Performance},
  volume={31},
  number={12},
  pages={9727--9739},
  year={2022},
  publisher={Springer}
}

@article{guillaume1904invar,
  author  = {Guillaume, Charles-Edouard},
  title   = {Invar and Its Applications},
  journal = {Nature},
  year     = {1904},
  volume   = {71},
  pages    = {134--139},
  doi      = {10.1038/071134a0}
}

@misc{carpenter_invar36,
  author       = {{Carpenter Technology Corporation}},
  title        = {INVAR 36 Datasheet},
  year         = {2024},
  url = {https://www.carpentertechnology.com/hubfs/7407324/Material%20Saftey%20Data%20Sheets/Invar%2036.pdf}
}

\end{document}